\documentclass{svproc}
\usepackage{graphicx}
\usepackage{amsmath}
\usepackage{booktabs}
\usepackage{times}
\usepackage{multirow}
\usepackage{amssymb}
\usepackage{xcolor}
\usepackage[misc]{ifsym}
\usepackage{caption}
\usepackage{subcaption}
\DeclareMathOperator*{\argmax}{arg\,max}

\begin{document}
\title{The benefits of coordination in (over)adaptive virtual teams}

\titlerunning{On the moderating effects of individual learning and team composition}
%
\author{Darío Blanco-Fernández\textsuperscript{(\Letter)}
\and
Stephan Leitner
\and
Alexandra Rausch
}
\authorrunning{D. Blanco-Fernández et al.}
%
\institute{University of Klagenfurt, Klagenfurt, 9020, Austria\\ \email{\{dario.blanco, stephan.leitner, alexandra.rausch\}@aau.at}}
\maketitle              
\begin{abstract}
The emergence of new organizational forms--such as virtual teams--has brought forward some challenges for teams. 
One of the most relevant challenges is coordinating the decisions of team members who work from different time zones. Intuition suggests that task performance should improve if the team members' decisions are coordinated. However, previous research suggests that the effect of coordination on task performance is ambiguous. Specifically, the effect of coordination on task performance depends on aspects such as the team members' learning and the changes in team composition over time. This paper aims to understand how individual learning and team composition moderate the relationship between coordination and task performance. We implement an agent-based modeling approach based on the \textit{NK}-framework to fulfill our research objective. Our results suggest that both factors have moderating effects. Specifically, we find that excessively increasing individual learning is harmful for the task performance of fully autonomous teams, but less detrimental for teams that coordinate their decisions. In addition, we find that teams that coordinate their decisions benefit from changing their composition in the short-term, but fully autonomous teams do not. In conclusion, teams that coordinate their decisions benefit more from individual learning and dynamic composition than teams that do not coordinate. Nevertheless, we should note that the existence of moderating effects does not imply that coordination improves task performance. Whether coordination improves task performance depends on the interdependencies between the team members' decisions.

\end{abstract}

\section{Introduction}\label{sec:intro}
Recent events have substantially changed the way organizations structure and deal with tasks. For example, the COVID-19 pandemic, the development of communication technologies, and the outsourcing of tasks in online platforms have increased teleworking arrangements \cite{BelzuneguiEraso2020,Kuhn2017,Squicciarini2011}. New forms of collaboration--such as virtual teams--have emerged in response to these changes in the workplace \cite{Puranam2014,Saunders2006,Squicciarini2011}. 

Virtual teams are \textit{"groups of people with a common purpose who carry out interdependent tasks across locations and time, using technology to communicate much more than they use face-to-face meetings"} \cite{Cramton2001}. Virtual teams are often self-organized, and decision-making is usually decentralized \cite{Puranam2014}. 
Decentralization and dispersion make it difficult for virtual team members to communicate their intended decisions to the rest of the team \cite{Cramton2001,Foster2001,Puranam2014,Saunders2006}. This might result in challenges in the coordination of the virtual teams' decisions.

Although intuition suggests that coordinating the team's decisions should increase task performance \cite{Edmondson2007}, previous results indicate that the effect of coordination on task performance is not straightforward \cite{Siggelkow2005}. Coordination does not unfold positive effects if tasks are simple; but is beneficial for sufficiently complex tasks \cite{Siggelkow2005,Wall2018}. In addition, coordination is beneficial when teams have access to a large set of solutions to the task they face \cite{Siggelkow2005}. Team members might generate this large set of solutions by \textit{learning} about the task and gradually adapting the knowledge to the task's requirements \cite{Blanco-Fernandez2022,Simon1991}. According to previous research, coordination allows teams to take full advantage of their members' learning \cite{Edmondson2001,Edmondson2007}.

Individual learning is not the only mechanism that teams use to develop new solutions to the task they face. Teams might change their composition over time to integrate new members that bring knowledge previously unavailable to the team \cite{Blanco-Fernandez2022b,Blanco-Fernandez2022,Simon1991}. Therefore, we expect that teams with a \textit{dynamic composition} identify new solutions faster than stable teams \cite{Blanco-Fernandez2022,Chang2005,Edmondson2001}. However, prior research on the effects of coordination on task performance usually focuses on teams which do not change their composition \cite{Siggelkow2005,Wall2018}. Consequently, previous research ignores the effects of dynamic team composition in the relationship between coordination and task performance. Our paper aims to fill this research gap.

The focus of this paper lies on self-organized teams--such as virtual teams--that solve complex tasks. Our objective is to understand how the effect of coordination on task performance is moderated by \textit{(i)} individual learning and \textit{(ii)} team composition. To achieve this research objective, we build on previous research by introducing coordination between the team members’ decisions \cite{Blanco-Fernandez2022a,Blanco-Fernandez2022b,Blanco-Fernandez2022}. We contribute to the literature by showing how coordination allows teams to fully grasp the benefits of individual learning and dynamic composition.

The remainder of this paper is organized as follows: Section \ref{sec:model} provides details on the model. Section \ref{sec:results} provides a description and discussion of the main results. Finally, Sec. \ref{sec:conclusion} concludes the paper.
\section{The Model}
\label{sec:model}
\begin{figure}[h]
    \centering
    \includegraphics[width=1\textwidth]{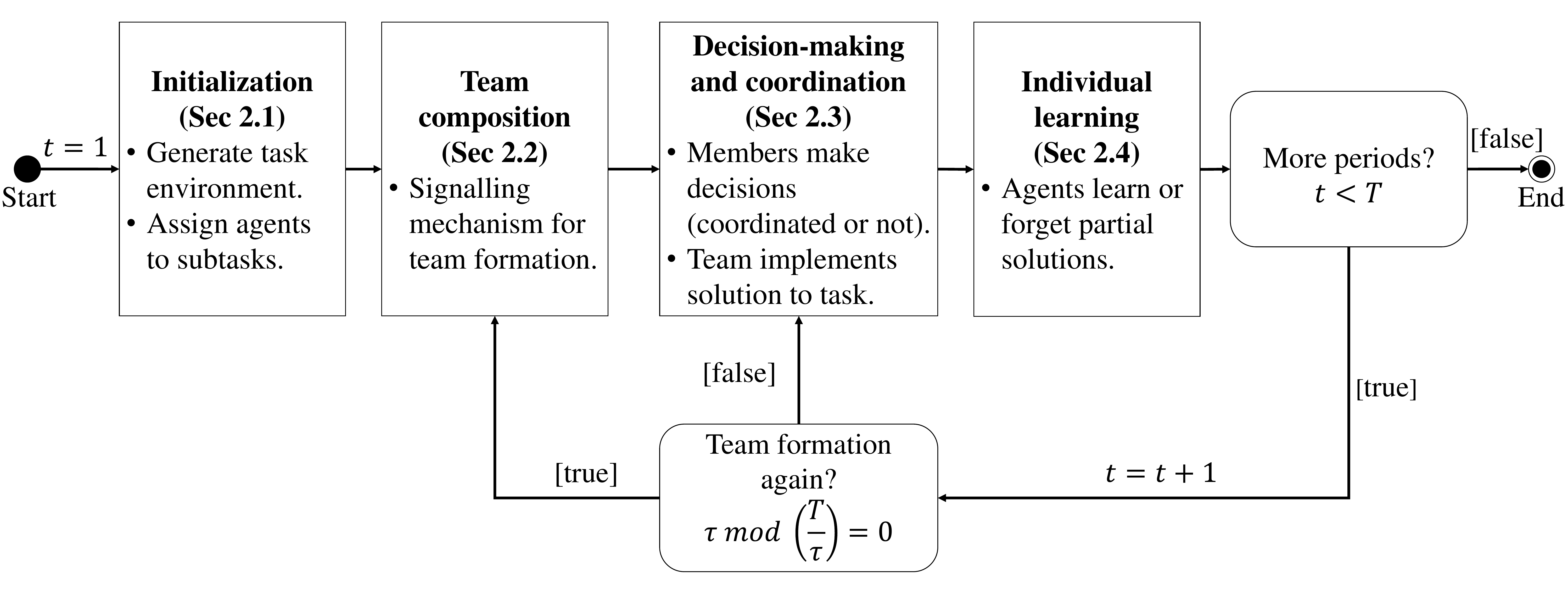}
    \caption{Temporal sequence of the model.}
    \label{fig:sequence}
\end{figure}

We implement an agent-based model based on the \textit{NK}-framework \cite{Levinthal1997}.\footnote{The model has been implemented in Python 3.7.4.} The model consists of four building blocks: The task environment and the agents (see Sec. \ref{sec:environment}), team formation (see Sec. \ref{sec:team}), decision-making and coordination (see Sec. \ref{sec:decision}), and individual learning (see Sec. \ref{sec:learning}). The four building blocks correspond to the sequence of events of the model, which we illustrate in Fig. \ref{fig:sequence}.

\subsection{Initialization}\label{sec:environment}
\subsubsection{Task environment}\label{sec:task}
The complex task that teams face consists of $N=12$ binary interdependent decisions. We divide the $N$-dimensional complex task into $M=3$ subtasks of equal length $S=N/M=4$. We denote each subtask by a vector $\mathbf{d}_m=(d_{S\cdot(m-1)+1},\dots, d_{S\cdot m})$ and the complex task by the vector $\mathbf{d}=(d_{1,}\dots, d_{N})$.\footnote{It follows that $\mathbf{d}_1 \frown \dots \frown \mathbf{d}_M = \mathbf{d}$, where $^\frown$ is the concatenation of each subtask.}

Each decision $d_{n} \in [0,1]$ contributes $c_{n}$ to task performance $C(\mathbf{d})$. This contribution depends on the decision itself and $K$ other decisions, so $c_{n}=f(d_{n}, d_{i_1}, \dots, d_{i_K})$, where $\{i_1, \dots, i_K \} \subseteq \{1, \dots, n-1, n+1, \dots, N \}$ and $0 \leq K \leq N-1$. We randomly generate contributions using a uniform distribution, $c_{n}\sim U(0,1)$. The overall task performance is the average of all contributions $C(\mathbf{d})=\frac{1}{N}\sum_{n=1}^{N}c_{n}$.

Each possible vector of $N=12$ binary values is a \textit{solution} to the complex task and has an associated performance. There are $2^{S}=16$ possible \textit{partial solutions} to each subtask and $2^{N}=4.096$ possible solutions to the complex task. The mapping of each solution to its associated performance is the \textit{performance landscape}. The team moves gradually on the performance landscape, following a steepest ascent hill-climbing search for new, better-performing solutions. 

The \textit{task complexity}--determined by $K$--partly influences the success of the search process. The higher is $K$, the more complex is the task, and the more rugged is the performance landscape \cite{Levinthal1997}. Several local maxima characterize a rugged performance landscape. Consequently, the higher the task complexity, the more likely it is for teams to get stuck at suboptimal solutions \cite{Levinthal1997}. Regarding task complexity, we consider two different scenarios: Low ($K=3$) and moderate complexity ($K=5$).

\begin{figure}[h]
    \centering
    \includegraphics[width=1\textwidth]{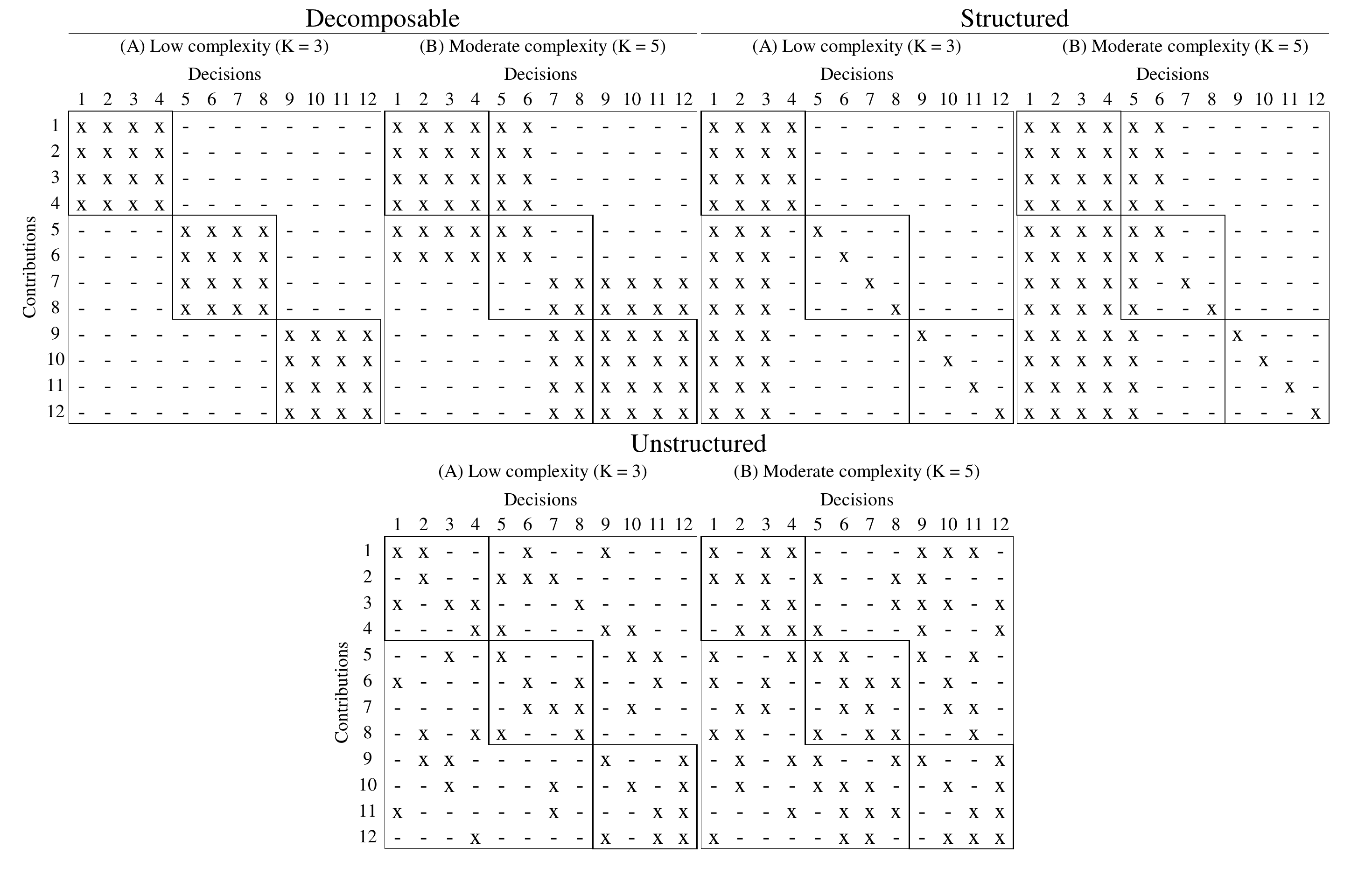}
    \caption{Interdependence matrices. Each matrix depicts which contributions (y-axes) depend on which decisions (x-axes). Interdependencies are indicated with an $X$. Since a contribution depends on its own decision, there is an $X$ in each element of the main diagonal. Solid lines indicate the subtasks.}
    \label{fig:matrices}
\end{figure}

The \textit{interdependence pattern} also affects the performance landscape's shape \cite{Rivkin2007}. The interdependence pattern reflects which contributions depend on which decisions. 
We consider three interdependence patterns, which we represent in Fig. \ref{fig:matrices}:

\begin{itemize}
    \item \textit{Decomposable}: Interdependencies are shaped in squares of size $K+1$. For $K=3$, the task is \textit{perfectly decomposable}, and there are no interdependencies between subtasks. For $K=5$, by contrast, there are interdependencies between subtasks.
    \item \textit{Structured}: The $K$ first decisions affect the remaining contributions. Thus, there is one subtask that heavily influences task performance.
    \item \textit{Unstructured}: Interdependencies between decisions are randomly allocated throughout the task, resulting in interdependencies between subtasks in all cases.
\end{itemize}

\subsubsection{Agents}\label{sec:agents}
To solve the complex task, a team is formed by choosing $M=3$ members out of a population of $P=30$ agents. These agents are heterogeneous and have \textit{limited capabilities} concerning the complex task. We limit the agents' capabilities in two ways. First, each agent can only solve one subtask $m$. Second, in the first period, we endow each agent with just one random partial solution to subtask $m$. Agents must solve the entire complex task to experience positive utility.

Agents are myopic as they aim to optimize just their immediate utility. Only team members experience positive utility.\footnote{Agents who do not join the team in one period get utility equal to $0$.} The utility function of an agent assigned to subtask $m$ is the weighted sum of their own performance contributions $C(\mathbf{d}_{mt})$ and the performance contributions of the \textit{residual decisions} $C(\mathbf{d}_{rt})$, where $r=\{1,\dots,M\}\in \mathbb{N}$ and $r \neq m$. We denote the residual decisions by $\mathbf{D}_{mt} = (\mathbf{d}_{1t}, \dots, \mathbf{d}_{\{m-1\}t}, \mathbf{d}_{\{m+1\}t},\dots, \mathbf{d}_{Mt})$. Agent $m$'s utility is calculated using Eq. \ref{eq:utility}:


\begin{equation}\label{eq:utility}
    U(\mathbf{d}_{mt}, \mathbf{D}_{mt}) = \frac{1}{2} \cdot \left( C(\mathbf{d}_{mt}) + \cdot \frac{1}{M-1} \sum_{\substack{{r=1}\\{r\neq m}}}^{M} C(\mathbf{d}_{rt})\right) ~.
\end{equation}

\subsection{Team composition}\label{sec:team}
Agents always have an incentive to participate in the team since it is the only way to experience positive utility. We assume that agents are fully aware of how team formation works and that they do not cheat. Additionally, we omit communication between agents during team formation. These assumption assure that agents do not behave strategically or form beliefs about other agents. Finally, we assume that one agent per subtask is sufficient to solve the complex task. Consequently, $M=3$ members form the team.

The objective of team formation is to assure that the current team members are the best-available agents for solving the task at any given period. Team formation works as follows. The set of solutions agent $m$ knows is ${\mathbf{S}}_{m}, =\left(\hat{\mathbf{d}}_{m1},\dots,\hat{\mathbf{d}}_{mI}\right)$ where $\hat{\mathbf{d}}_{mi}$ is a solution to subtask $\mathbf{d}_m$, $i=\{1,\dots,I\}\in \mathbb{N}$ and $1 \leq I \leq 2^S$. Each agent estimates the utility for each solution they know i.e., $\forall \hat{\mathbf{d}}_{mi} \in \mathbf{S}_{mt}$. Since we omit communication, agents use the residual decisions from the previous period $\mathbf{D}_{m\{t-1\}}$ as a basis for their estimations. Agent $m$'s \textit{estimated utility} is then: 

\begin{equation}\label{eq:estutility}
    EU(\mathbf{d}_{mt}, \mathbf{D}_{m\{t-1\}}) = \frac{1}{2} \cdot \left( C(\mathbf{d}_{mt}) + \cdot \frac{1}{M-1} \sum_{\substack{{r=1}\\{r\neq m}}}^{M} C(\mathbf{d}_{r\{t-1\}})\right) + e ~;
\end{equation}

\noindent where $e$ is an error term which follows a normal distribution $e\sim N(0,0.01)$. This error term reflects the mistakes that team members might make when estimating the effects of their decisions \cite{Leitner2015,Wall2020}.

After the estimation, each agent signals the highest estimated utility $U(\hat{\mathbf{d}}^{*}_{mt}, \mathbf{D}_{m\{t-1\}})$, where $\hat{\mathbf{d}}^{*}_{mt} := \argmax_{\mathbf{d}^\prime \in \mathbf{S}_{mt}} ~ U(\mathbf{d}^\prime, \mathbf{D}_{m\{t-1\}})$ is the solution that maximizes agent $m$'s estimated utility at time $t$. The agent who signals the highest estimated utility for each subtask $m$ becomes a team member. 

The agents form the first team iteration in the first period. Afterwards, team formation is repeated every $\tau$ periods. The higher (lower) $\tau$, the less (more) frequently a team changes its composition. We study three different scenarios for $\tau$:

\begin{itemize}
    \item Teams with a \textit{long-term composition} do not change their composition over time. We denote this scenario by $\tau=\emptyset$.
    \item Teams with a \textit{medium-term composition} change their composition every $\tau=10$ periods.
    \item Teams with a \textit{short-term composition} change their composition at every period, i.e., $\tau=1$.
\end{itemize}

\subsection{Decision-making and coordination}\label{sec:decision}
The $M=3$ team members choose a team solution to the complex task at every period. In the benchmark scenario, members of a \textit{fully autonomous} team make their choices independently and simultaneously \cite{Siggelkow2005,Wall2018}. They calculate the estimated utility for every partial solution they know, i.e. $\forall \hat{\mathbf{d}}_{mi} \in \mathbf{S}_{mt}$, following Eq. \ref{eq:estutility}. Each member's choice is $\hat{\mathbf{d}}^{*}_{mt}$, i.e., the partial solution associated with the highest estimated utility. Finally, the concatenation of all member's choices is the team solution for the current period $\mathbf{d}_{t} :=\hat{\mathbf{d}}^{*}_{1t}\frown \dots \frown \hat{\mathbf{d}}^{*}_{Mt}$.

We contrast this benchmark scenario with an scenario in which the team coordinates its decisions. The coordination mechanism is based on the \textit{liaison} organizational archetype described in \cite{Siggelkow2005}. We assume that all agents know how the coordination mechanism works. Initially, each team member ranks all partial solutions they know $ \hat{\mathbf{d}}_{mi} \in \mathbf{S}_{mt}$ regarding their estimated utility (see Eq. \ref{eq:estutility}). Then each member chooses the two highest partial solutions $\hat{\mathbf{d}}^{(1)}_{mt}$ and  $\hat{\mathbf{d}}^{(2)}_{mt}$, where the solution with the highest expected utility is ranked first and the solution with the second highest expected utility is ranked second. The team members bring these partial solutions to a coordination session. 

Two candidate solutions are constructed in order, first by concatenating the preferred choices and then the second-preferred choices so  $\mathbf{d}^{(j)}_{t} :=\hat{\mathbf{d}}^{(j)}_{1t}\frown \dots \frown \hat{\mathbf{d}}^{(j)}_{Mt}$ where $\hat{\mathbf{d}}^{(j)}_{mt}$ is agent $m$'s $j^{th}$ preferred choice. Each team member sequentially evaluates the two candidate solutions $\mathbf{d}^{(j)}_{t}$ regarding their estimated utility. If the estimated utility of a candidate solution is higher than the last achieved utility, the team member accepts the candidate solution, i.e., they accept the solution if $EU_m(\mathbf{d}^{(j)}_{t} )>U_m(\mathbf{d}_{t-1})$. Otherwise, they veto it. The veto from one member is enough to reject the candidate solution. If all members accept a candidate solution, it is chosen as the team solution for the current period, so $\mathbf{d}_{t}=\mathbf{d}^{(j)}_{t}$. Conversely, the team solution remains constant from the previous period if members veto both candidate solutions, so $\mathbf{d}_{t}=\mathbf{d}_{t-1}$.

\subsection{Individual learning}\label{sec:learning}
Agents overcome their limited capabilities by \textit{learning} about subtask $m$. Specifically, agents learn by exploring the solution space and changing their set of partial solutions ${\mathbf{S}}_{m}$ over time \cite{Levinthal1997}. Learning occurs at the end of each period and consists of two separate mechanisms. First, with probability $\mathbb{P}$, agents might \textit{discover} a partial solution. The partial solution they discover differs only in the value of one decision from any currently-known partial solution. Second, with the same probability $\mathbb{P}$, agents might \textit{forget} a partial solution that is not utility-maximizing at the current period. We study probabilities of learning between $\mathbb{P}=0$ and $\mathbb{P}=1$ in intervals of $0.1$.

\subsection{Parameters and performance measures}\label{sec:variables}
Our research comprises 396 different scenarios. Each scenario consists of 1,500 simulation rounds of $T=200$ periods each. We summarize the main parameters of the model and their values in Tab. \ref{tab:variables}.

\begin{table}
\caption{Parameters}
\label{tab:variables}
\renewcommand{\arraystretch}{1.2}
\begin{tabular}{llll}
\\ \hline
Type                                  & Variables              & Notation                    & Values                         \\ \hline
\multirow{6}{*}{Independent variables} & Task complexity        & $K$                         & \{3, 5\}                       \\
                                      & Interdependence structure & \textit{Matrix} & See Fig. \ref{fig:matrices}     \\
                                      & Team composition      & $\tau$                      & \{$\emptyset, 1, 10$\}         \\
                                      & Learning probability   & $\mathbb{P}$                & $\{0:0.1:1\}$ \\
                                      & Time period             & $t$                         & $\{1:1:100\}$   \\
                                      & Coordination        & N/A                         & \textit{Fully autonomous, Coordination} \\\hline
Dependent variable                    & Task performance       & $C(\mathbf{d_t})$           & $[0,1]$                     \\ \hline
\multirow{5}{*}{Other parameters}       
                                      & Number of decisions    & $N$                         & 12                             \\
                                      & Population of agents   & $P$                         & 30                             \\
                                      & Number of subtasks     & $M$                         & 3                              \\
                                      & Number of simulations  & $\Phi$                         & 1,500                        \\
                                      & Error term             & $e$ & $e\sim N(0,0.01)$ \\
\hline
\end{tabular}%
\end{table}

We normalize the observed task performance at each period $C(\mathbf{d_{t}})$ by the maximum achievable performance at each simulation round $C^{\ast}$. Normalization assures that we can compare different scenarios in terms of task performance.

We use the normalized performances to train regression tree models using task performance as the dependent variable and the independent variables of Tab. \ref{tab:variables}. We then compute partial dependencies between task performance and the moderating factors, i.e., individual learning and team composition. To calculate partial dependencies, we first define $\mathbf{X}$ as the set of all independent variables. The set $\mathbf{X}$ is divided into two subsets. Subset $\mathbf{X}^s$ corresponds to the scope variable, i.e., individual learning or dynamic team composition. Subset $\mathbf{X}^c$ includes the remaining independent variables.\footnote{It follows that $\mathbf{X}^s \cup \mathbf{X}^c = \mathbf{X}$.} We compute the partial dependence of task performance on the moderating factor studied according to $f^s(\mathbf{X}^s)= E_c(f(\mathbf{X}^s,\mathbf{X}^c)) \approx \frac{1}{V}\sum_{i=1}^{V} f(\mathbf{X}^s,\mathbf{X}_{(i)}^c)$, where $V$ is the number of independent variables in $\mathbf{X}^c$ and $\mathbf{X}_{(i)}^c$ corresponds to each variable. We employ this method to understand the patterns related to our research objective \cite{patel2018}. 
\section{Results and Discussion}\label{sec:results}
According to prior research, coordination is more beneficial for teams that search extensively for new solutions \cite{Rivkin2003,Siggelkow2005}. Teams may acquire new solutions because \textit{(i)} their members learn about the complex task or \textit{(ii)} they change their composition \cite{Simon1991}. Some authors advocate that research on coordination should consider search processes at multiple levels \cite{Tosic2010}. Our research follows this suggestion, and aims to understand how individual learning ($\mathbb{P}$) and dynamic team composition ($\tau$) moderate the relationship between coordination and task performance. In the following subsections, we study the moderating effects of individual learning (Sec. \ref{subsec:resultslearning}) and team composition (Sec. \ref{subsec:resultsteam}) separately.

\subsection{Moderating effect of individual learning}\label{subsec:resultslearning}
\begin{figure}[h]
    \centering
    \begin{subfigure}{0.345\linewidth}
        \captionsetup{justification=centering}
        \caption{Low complexity \\ Decomposable}
        \label{rq1a}
        \includegraphics[width=\linewidth]{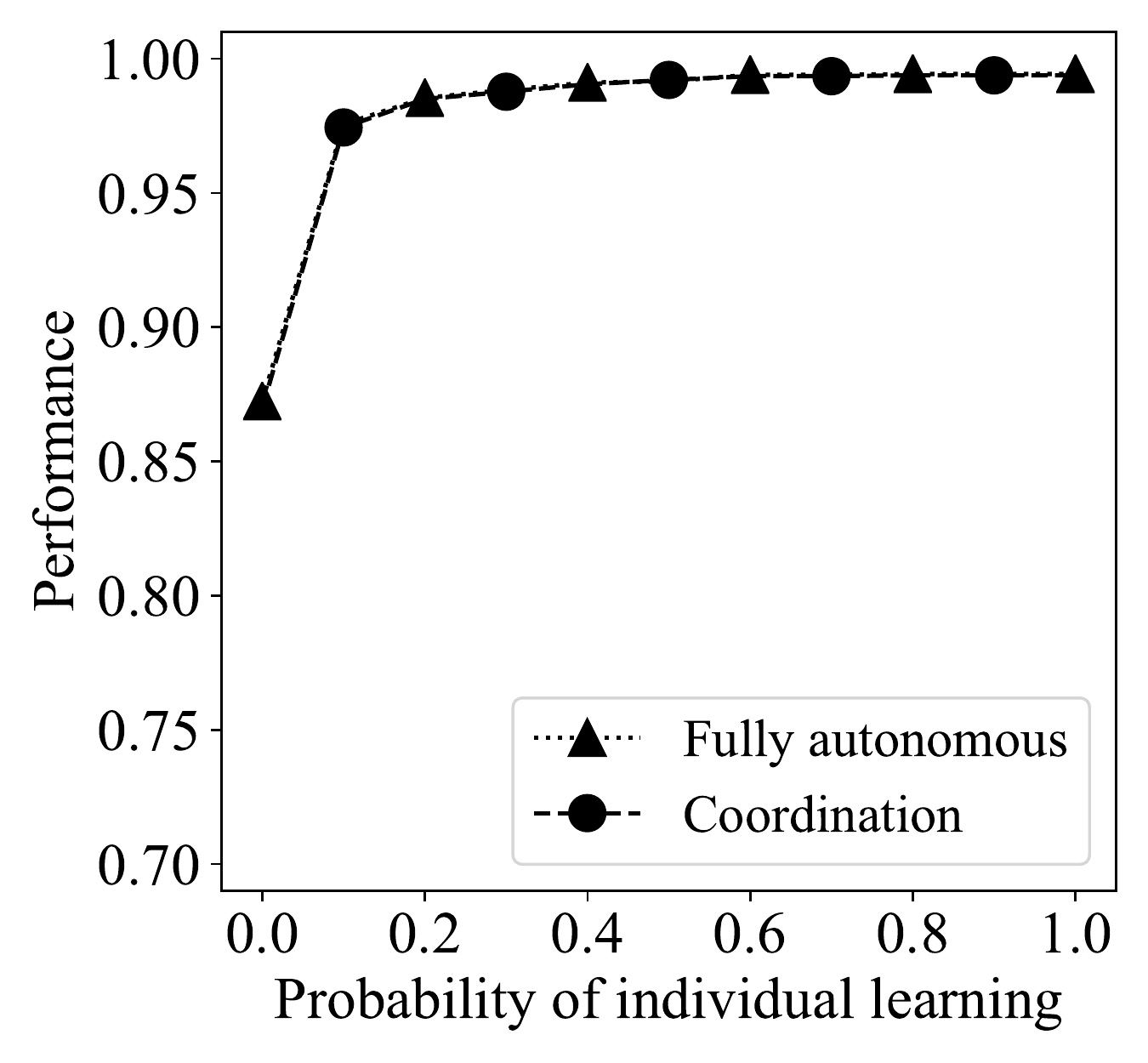}
    \end{subfigure}
    \begin{subfigure}{0.32\linewidth}
        \captionsetup{justification=centering}
        \caption{Low complexity \\ Structured}
        \label{rq1b}
        \includegraphics[width=\linewidth]{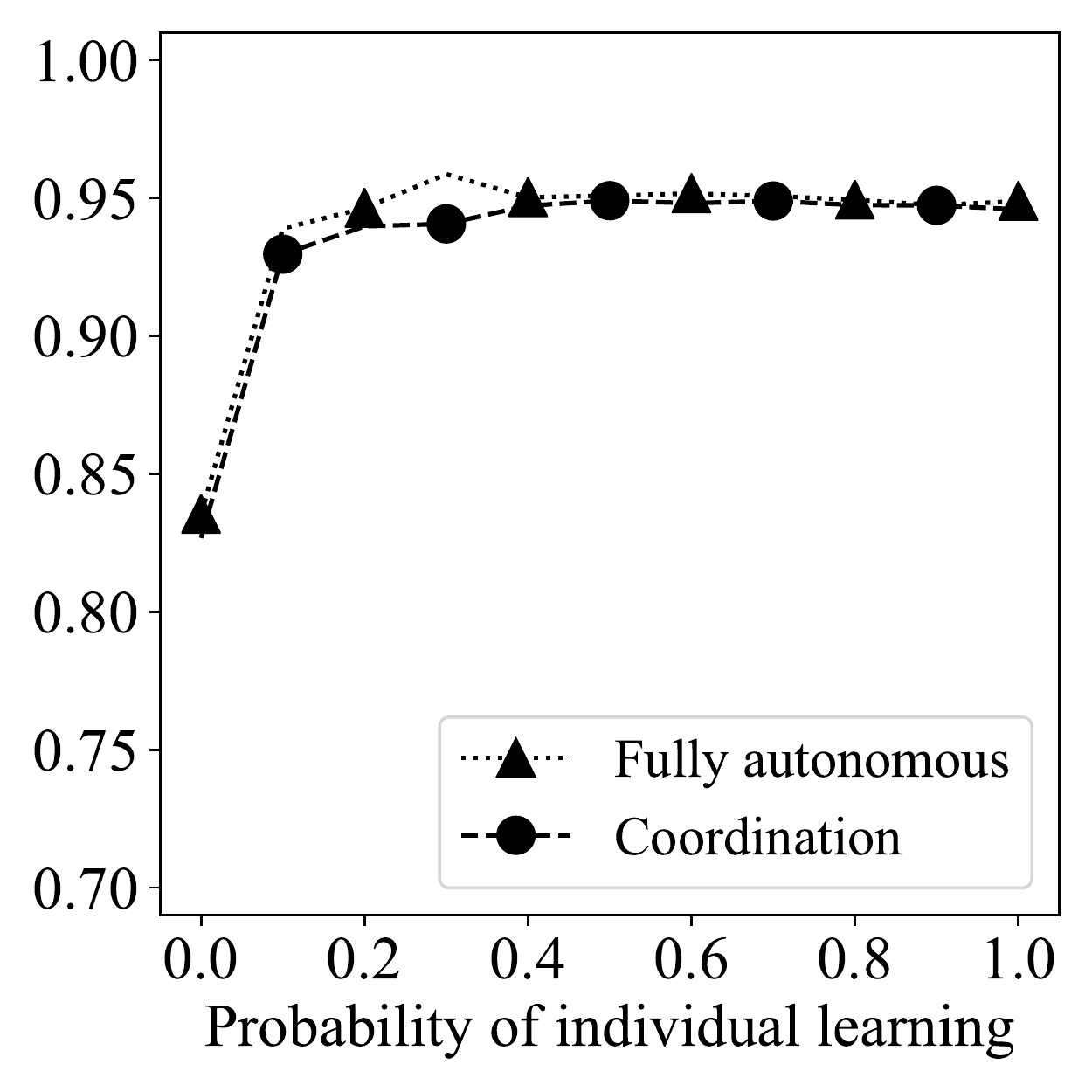}
    \end{subfigure}
    \begin{subfigure}{0.32\linewidth}
        \captionsetup{justification=centering}
        \caption{Low complexity \\ Unstructured}
        \label{rq1c}
        \includegraphics[width=\linewidth]{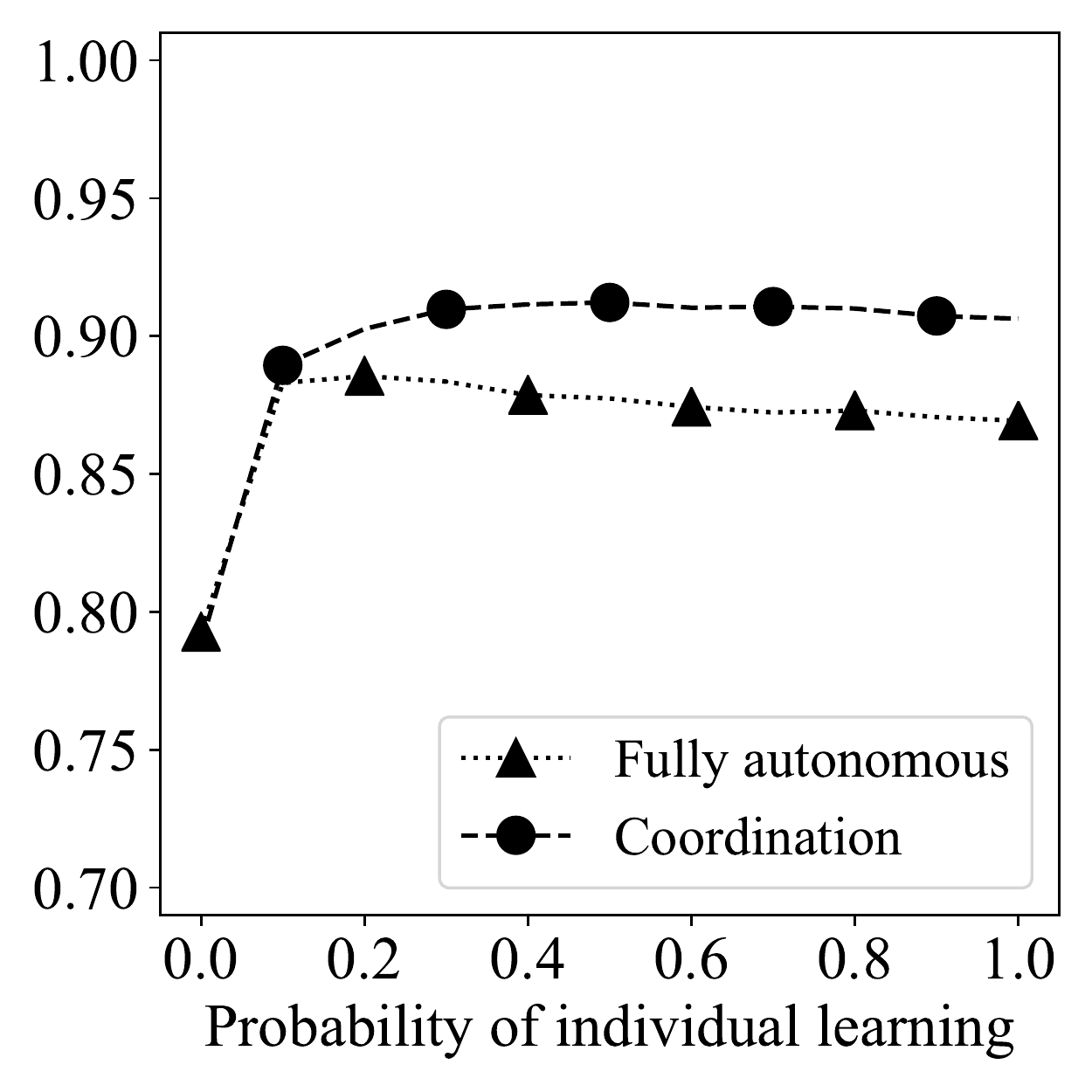}
    \end{subfigure}
    
    \begin{subfigure}{0.345\linewidth}
        \captionsetup{justification=centering}
        \caption{Moderate complexity \\ Decomposable}
        \label{rq1d}
        \includegraphics[width=\linewidth]{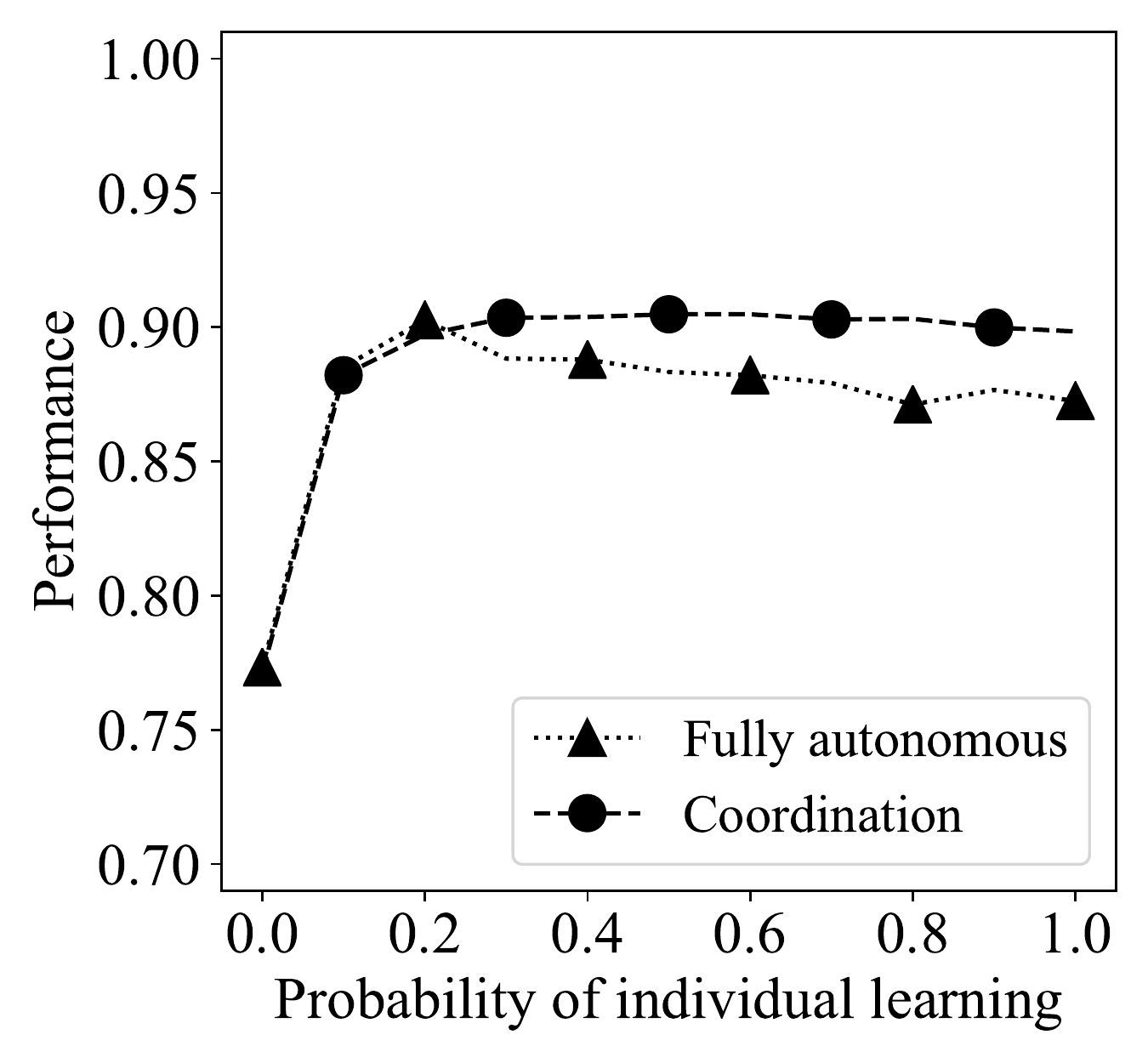}
    \end{subfigure}
     \begin{subfigure}{0.32\linewidth}
        \captionsetup{justification=centering}
        \caption{Moderate complexity \\ Structured}
        \label{rq1e}
        \includegraphics[width=\linewidth]{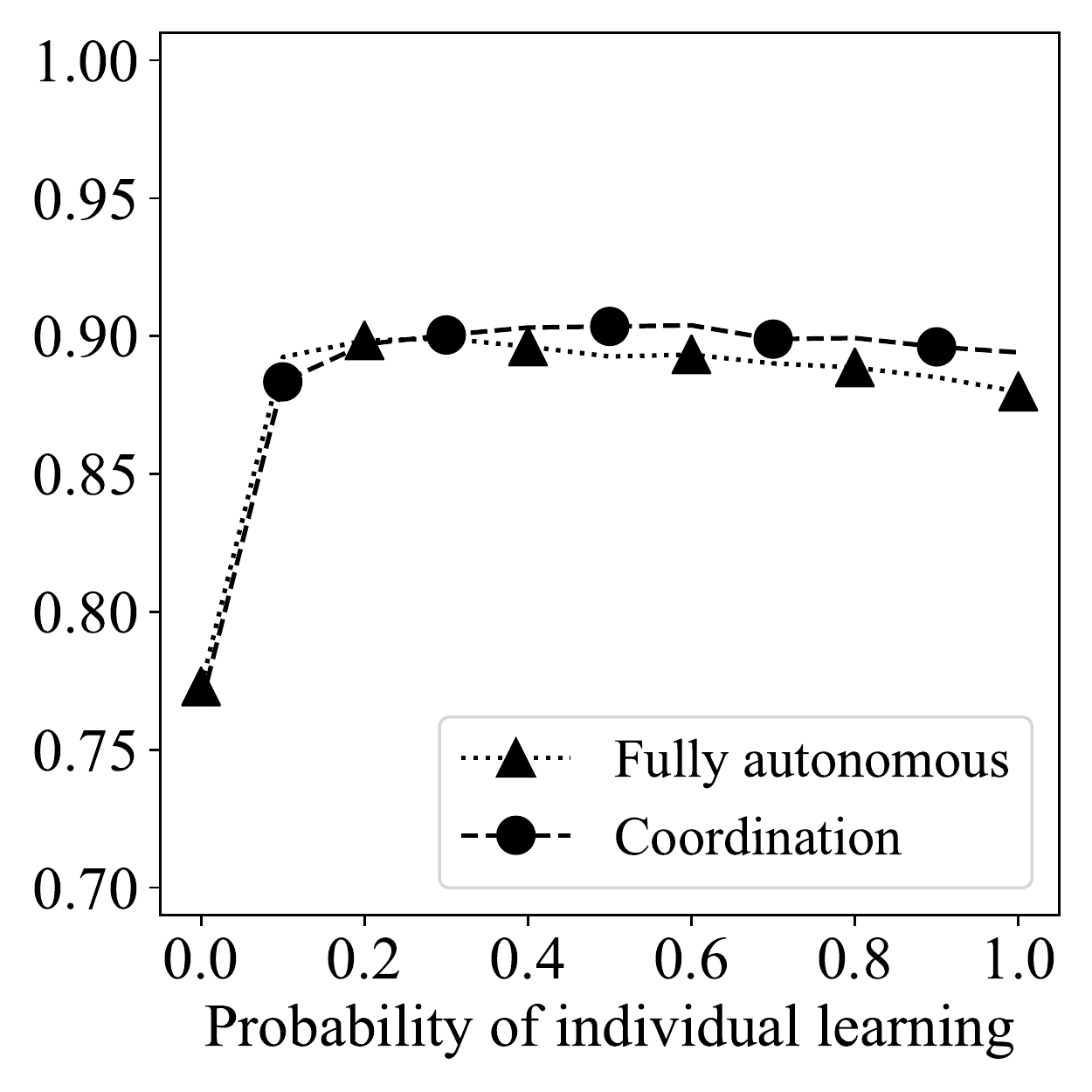}
    \end{subfigure}
    \begin{subfigure}{0.32\linewidth}
        \captionsetup{justification=centering}
        \caption{Moderate complexity \\ Unstructured}
        \label{rq1f}
        \includegraphics[width=\linewidth]{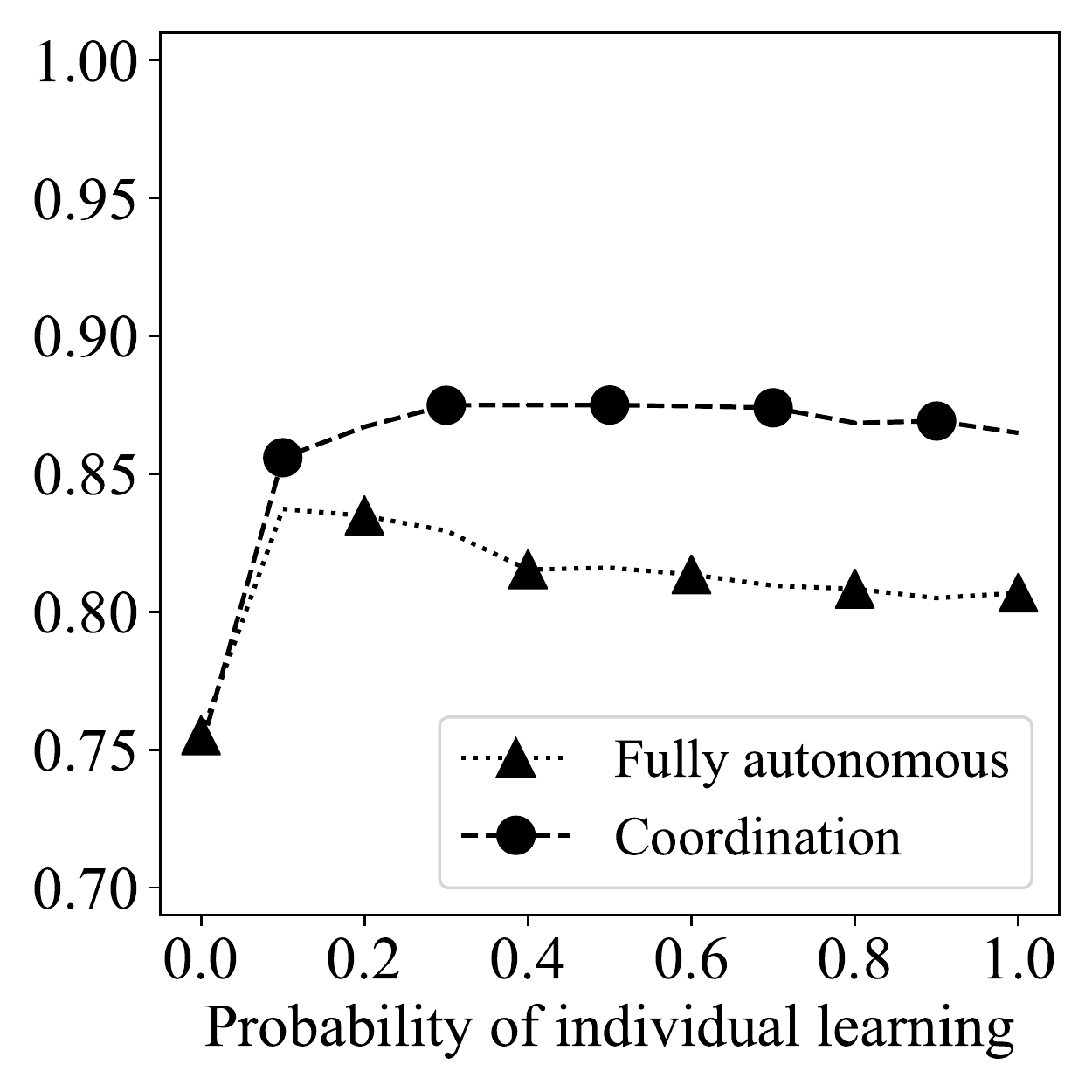}
    \end{subfigure}
    
    \captionsetup{justification=centering}
        \caption{Partial dependencies of task performance (y-axes) on the probability of individual learning (x-axes).}
    \label{fig:resultslearning}
\end{figure}

To study the moderating effect of individual learning, we calculate partial dependencies using the individual learning probability as the scope variable. We represent the results in Fig. \ref{fig:resultslearning}. Each plot shows the partial dependence of task performance on the learning probability for each level of complexity $K$ and for each interdependence structure. 

\noindent The moderating effect of individual learning depends on the level of complexity and the interdependence structure studied. There is no moderating effect of individual learning for decomposable and structured tasks of low complexity (Fig. \ref{rq1a} and Fig. \ref{rq1b}). The effect of increasing the individual learning probability on task performance is remarkably similar for teams that coordinate their decisions and fully autonomous teams. Task performance reacts strongly to initial increases in the probability of individual learning. However, the task performance stabilizes for higher values of  $\mathbb{P}$. 

Individual learning has a moderating effect for unstructured tasks of low complexity (Fig. \ref{rq1c}) and moderately complex tasks (Fig. \ref{rq1d}-Fig. \ref{rq1f}). At relatively low levels of learning, the effect of increasing $\mathbb{P}$ is similar for teams that coordinate their decisions and fully autonomous teams. Task performance increases, albeit the marginal positive effect decreases with each increase in $\mathbb{P}$. Eventually, there is a threshold value for the individual learning probability at which its effect on task performance turns negative. This negative effect is highly relevant for fully autonomous teams. In contrast, the decrease in task performance is barely notable in teams that coordinate their decisions. Consequently, the benefits of coordination increase with individual learning.

Our results show that agents learning more does not necessarily increase task performance, but might even harm it (Fig. \ref{rq1c}-Fig. \ref{rq1f}). Therefore, our results align with previous research, which highlights the negative impact that excessive individual learning might have on task performance \cite{Blanco-Fernandez2022a,Blanco-Fernandez2022}. In addition, we show that coordination reduces these adverse effects notably. The reason for this might lie in the evaluation of the team members' solutions. Coordination allows teams to evaluate its members' partial solutions more efficiently, increasing task performance \cite{Edmondson2007}. Consequently, teams that coordinate their decisions can grasp the full benefits of individual learning, fully autonomous teams cannot \cite{Edmondson2001,Edmondson2007}.

\subsection{Moderating effect of team composition}\label{subsec:resultsteam}
\begin{figure}[h]
    \centering
    \begin{subfigure}{0.345\linewidth}
        \captionsetup{justification=centering}
        \caption{Low complexity \\ Decomposable}
        \label{rq2a}
        \includegraphics[width=\linewidth]{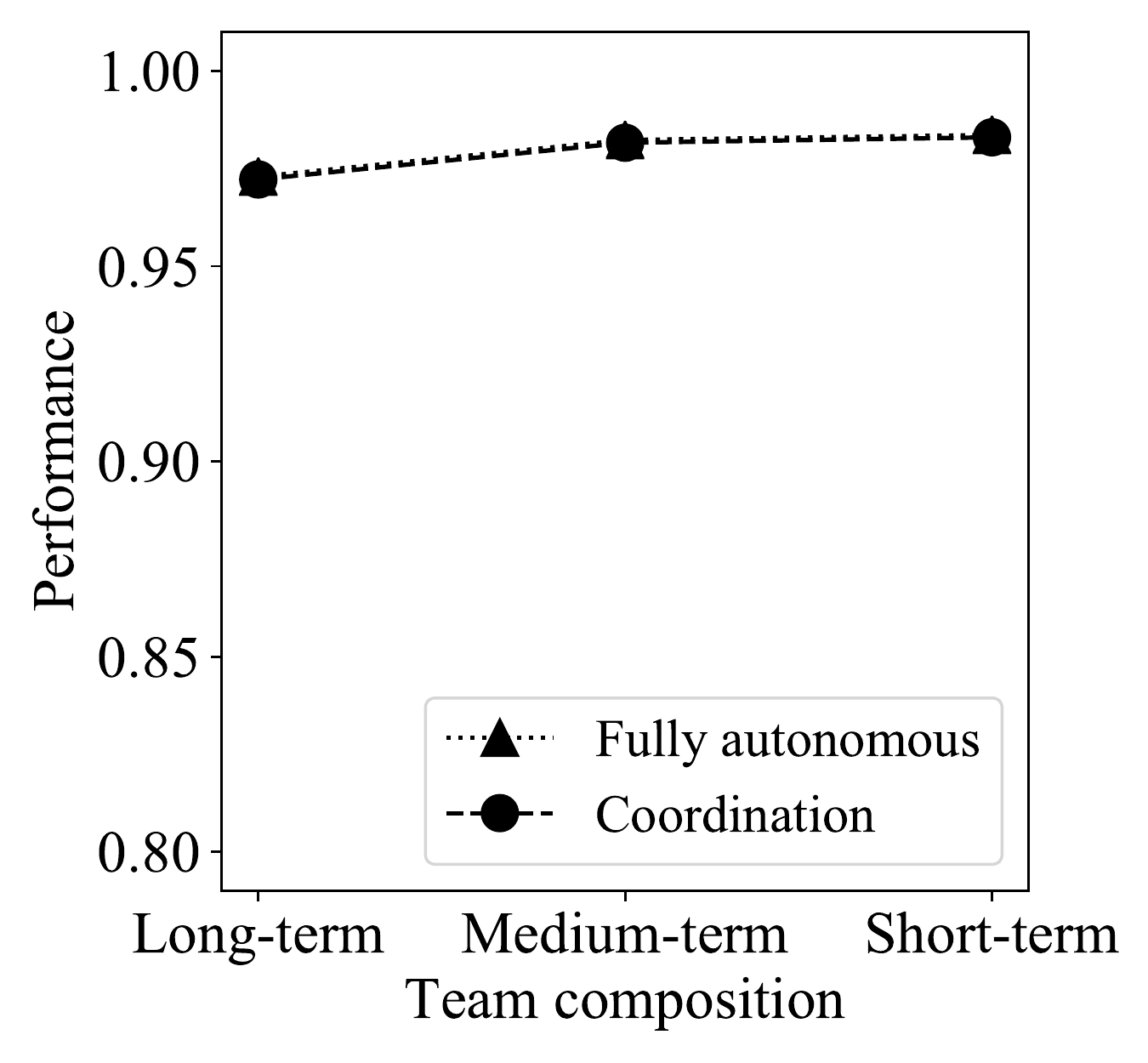}
    \end{subfigure}
    \begin{subfigure}{0.32\linewidth}
        \captionsetup{justification=centering}
        \caption{Low complexity \\ Structured}
        \label{rq2b}
        \includegraphics[width=\linewidth]{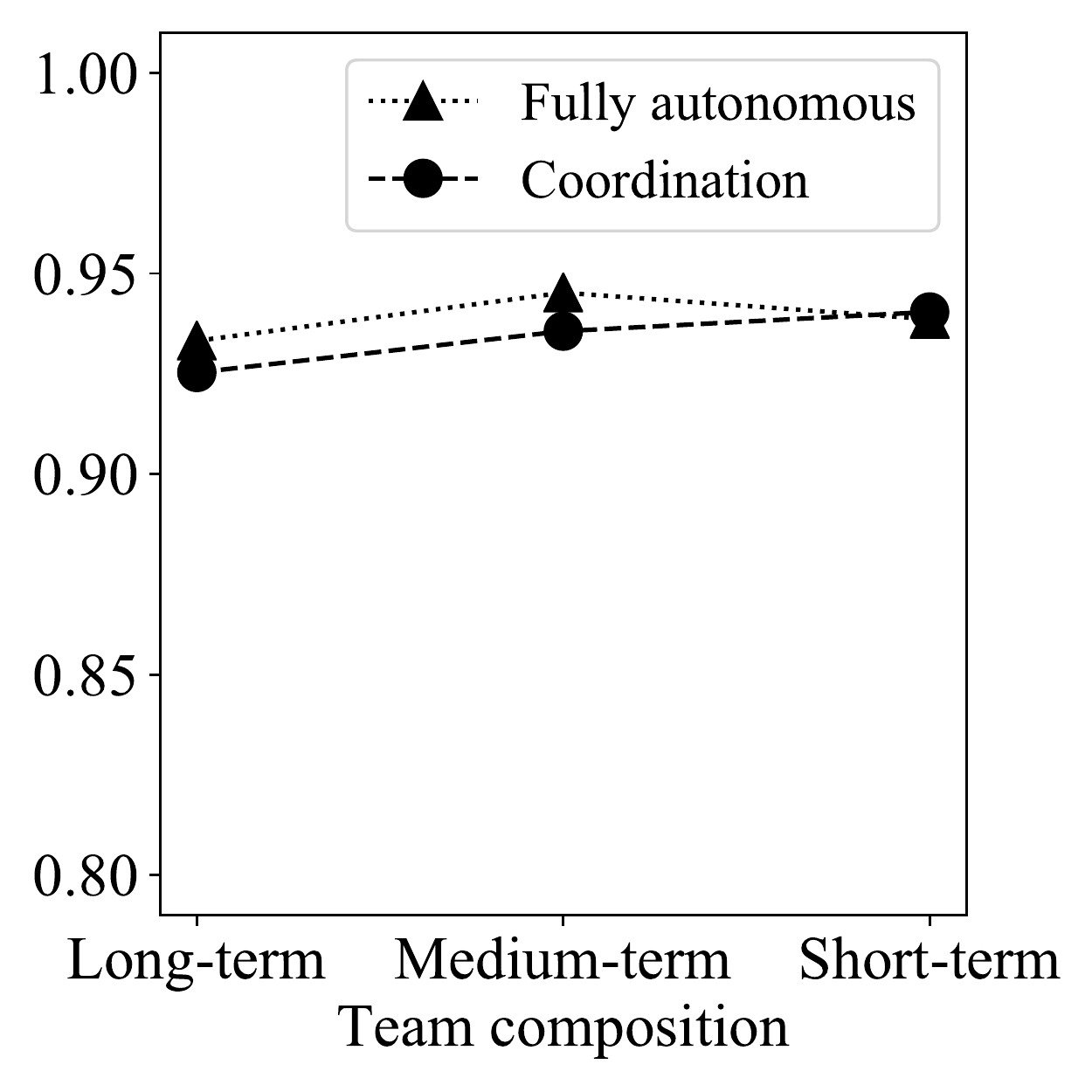}
    \end{subfigure}
    \begin{subfigure}{0.32\linewidth}
        \captionsetup{justification=centering}
        \caption{Low complexity \\ Unstructured}
        \label{rq2c}
        \includegraphics[width=\linewidth]{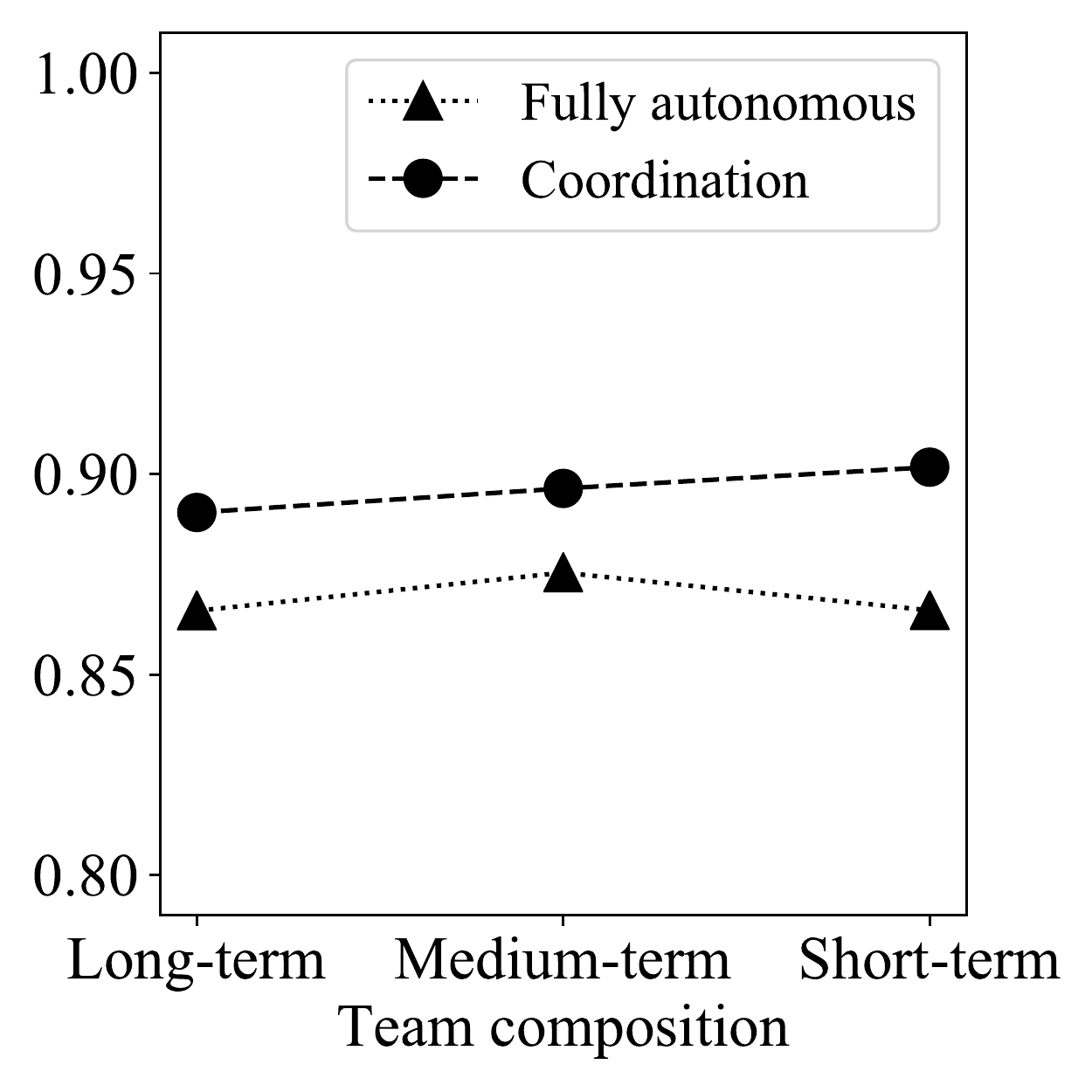}
    \end{subfigure}
    
    \begin{subfigure}{0.345\linewidth}
        \captionsetup{justification=centering}
        \caption{Moderate complexity \\ Decomposable}
        \label{rq2d}
        \includegraphics[width=\linewidth]{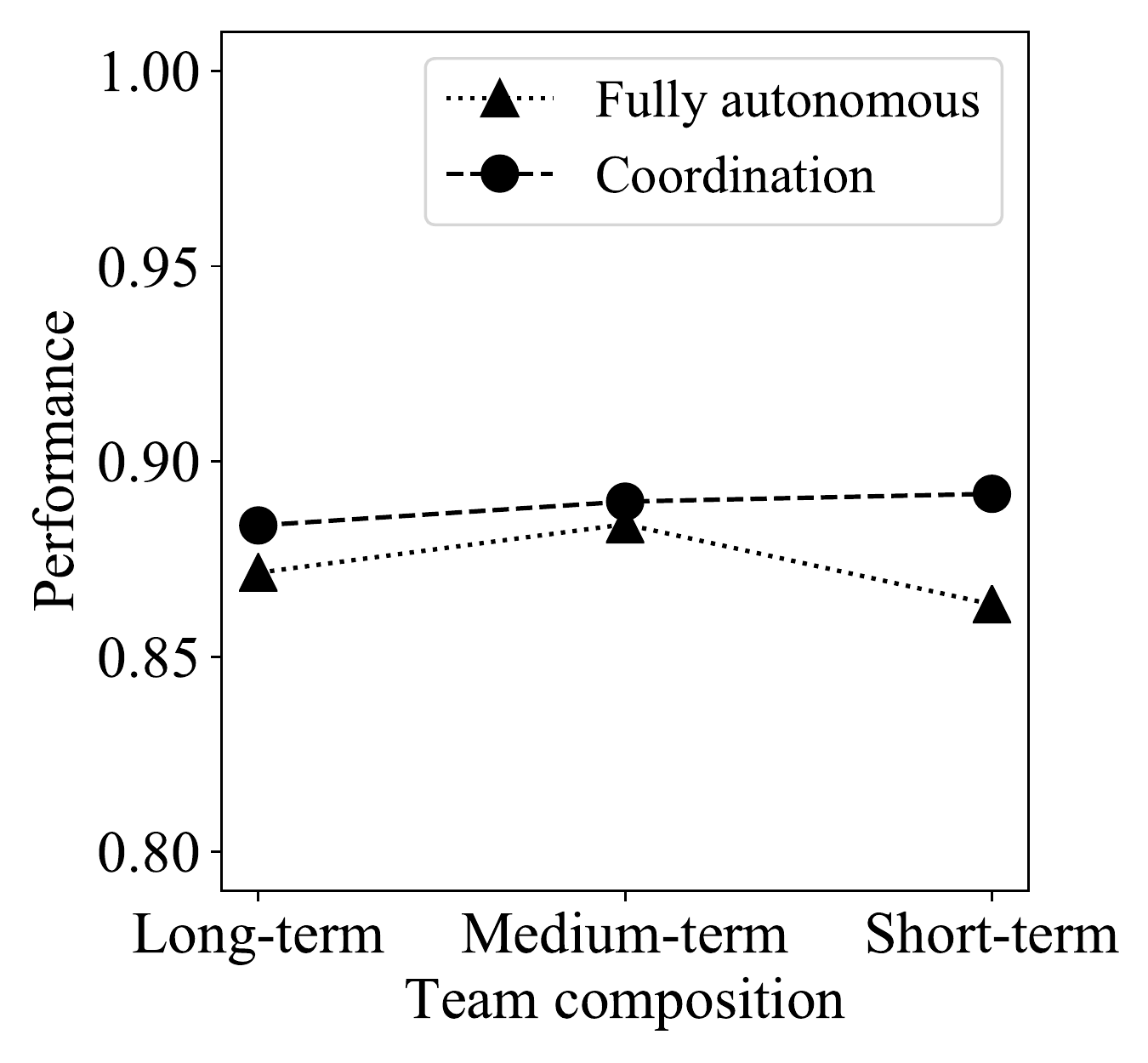}
    \end{subfigure}
     \begin{subfigure}{0.32\linewidth}
        \captionsetup{justification=centering}
        \caption{Moderate complexity \\ Structured}
        \label{rq2e}
        \includegraphics[width=\linewidth]{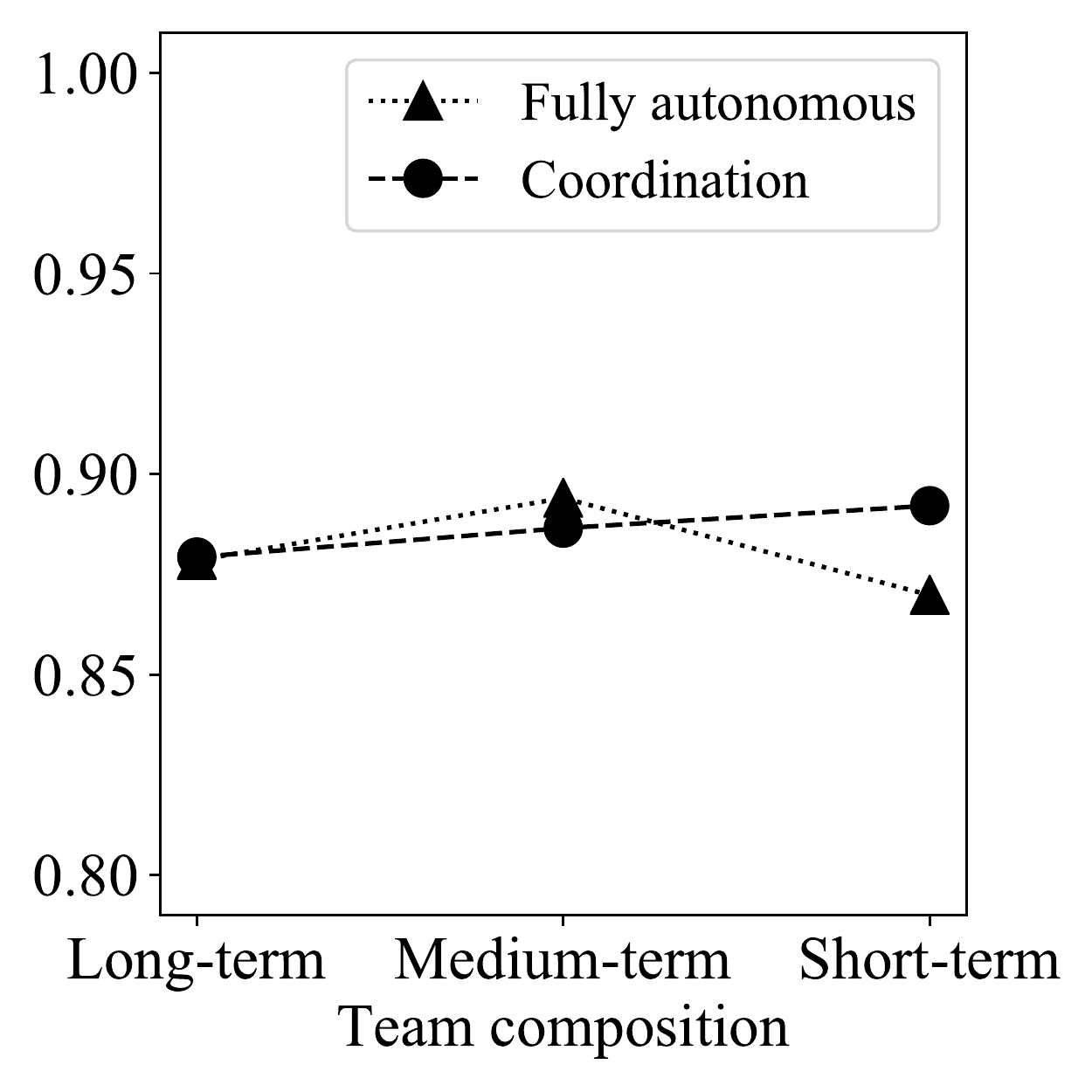}
    \end{subfigure}
    \begin{subfigure}{0.32\linewidth}
        \captionsetup{justification=centering}
        \caption{Moderate complexity \\ Unstructured}
        \label{rq2f}
        \includegraphics[width=\linewidth]{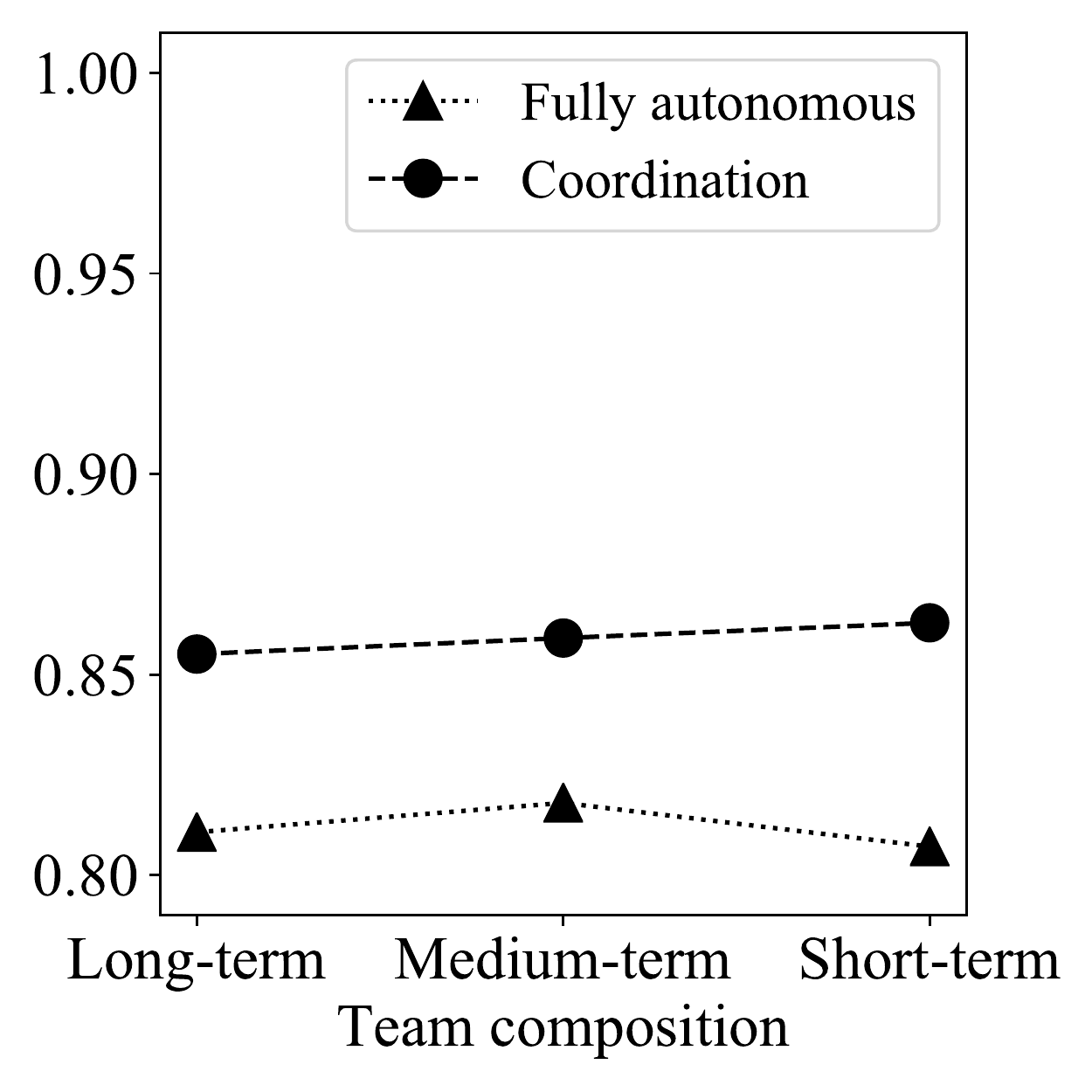}
    \end{subfigure}
    
    \captionsetup{justification=centering}
        \caption{Partial dependencies of task performance (y-axes) on team composition (x-axes).}
    \label{fig:resultsteam}
\end{figure}

We compute partial dependencies using team composition as the scope variable to study the second moderating effect. Each plot in Fig. \ref{fig:resultsteam} shows the partial dependence of task performance on team composition for each level of complexity $K$ and each interdependence structure considered.

In perfectly decomposable tasks, we find that team composition $\tau$ has the same impact on task performance regardless of coordination (see Fig. \ref{rq2a}). A dynamic team composition slightly improves task performance compared to a stable team composition, irrespective of the frequency of team formation.

For the remaining scenarios (Fig. \ref{rq2b}-Fig. \ref{rq2f}), there are different patterns depending on coordination. The task performance of teams that coordinate their decisions increases with the frequency of team formation. Consequently, teams that coordinate their decisions benefit more from changing their composition in the short-term. By contrast, fully autonomous teams only benefit from changing their composition in the medium-term. Changes in the short-term might unfold neutral (see Fig. \ref{rq2b}, Fig. \ref{rq2c}, and \ref{rq2f}) or negative effects (see Fig. \ref{rq2d} and Fig. \ref{rq2e}) on task performance.

Teams that change their composition find more solutions than stable teams \cite{Blanco-Fernandez2022a}, but they require coordination to translate this search process into improvements in task performance \cite{Siggelkow2005,Tannenbaum2012}. The lack of coordination between the team members' decisions makes fully autonomous teams unable to benefit from changing their composition in the short-term. In contrast, teams that coordinate their decisions benefit from more frequent changes in their composition (see Fig. \ref{rq2b}-Fig. \ref{rq2f}). Thus, our results align with the insights given in prior research \cite{Siggelkow2005,Tannenbaum2012}.

Whether this moderating effect results in coordination improving task performance depends on the interdependence structure. Coordination never improves task performance for structured tasks of low complexity (Fig. \ref{rq2b}). For decomposable and structured tasks of moderate complexity (Fig. \ref{rq2d} and Fig. \ref{rq2e}), coordination only improves task performance if teams have a short-term composition. Finally, coordination is always beneficial for unstructured tasks (Fig. \ref{rq2c} and Fig. \ref{rq2f}). Additionally, the positive effect of coordination in unstructured tasks is higher for a short-term team composition.

Our results align partially with the insights given by \cite{Tannenbaum2012}, who claim that dynamic teams should coordinate their decisions to improve task performance. We show that coordination is beneficial only in certain situations. The benefits of coordination grow with the frequency of changing composition and the interdependencies between the team members' decisions. According to prior research, teams might achieve indirect coordination by assigning the decisions in such a manner that the interdependencies between subtasks are minimized \cite{Rivkin2003}. If subtasks are less interdependent then there is less need for coordination, as the effect of the members' decisions on the remaining members' contributions diminishes \cite{Wall2018}. This indirect coordination is reflected in decomposable and structured tasks. Consequently, the impact on task performance of coordinating the team decisions is reduced \cite{Rivkin2003,Wall2018}.

\section{Summary and conclusions}\label{sec:conclusion}
This paper studies how individual learning and team composition moderate the relationship between coordination and task performance. We find moderating effects of both variables that are more relevant, the more interdependent are the team members' decisions. For example, coordination does not affect the performance of decomposable and structured tasks of low complexity. In contrast, coordination improves task performance for \textit{(i)} unstructured tasks, \textit{(ii)} for sufficiently high individual learning, or \textit{(iii)} for a short-term composition.

Regarding individual learning, our results suggest that the adverse effects of increasing individual learning excessively are reduced if teams coordinate their decisions. Regarding team composition, we find that fully autonomous teams only benefit from changing their composition in the medium-term. In contrast, teams that coordinate their decisions benefit more from changing their composition in the short term. 

Our research has some limitations. We only test the effect of one coordination mechanism. More coordination mechanisms could be added to extend this research, as in \cite{Siggelkow2005}. Additionally, prior research suggests that individual learning and team composition interact. \cite{Blanco-Fernandez2022}. Future research can study the joint moderating effect of individual learning and team composition on the relationship between coordination and task performance. Finally, researchers have suggested other aspects that might affect the relationship between team composition and coordination. In particular, researchers cite team identity as a relevant factor for coordination in dynamic teams \cite{Kogut1996,Saunders2006}. Future extensions of our research could consider this aspect.

%
%
%
\bibliographystyle{splncs04}
\bibliography{mybibliography}

\end{document}